# Entangled Quantum State of Magnetic Dipoles

S. Ghosh[1], T.F. Rosenbaum[1], G. Aeppli[2], and S.N. Coppersmith[3]

[1]*James Franck Institute and Department of Physics, University of Chicago, Chicago, IL 60637*

[2]*London Centre for Nanotechnology and Department of Physics and Astronomy, UCL, London, WC1E 6BT, UK and NEC Laboratories, 4 Independence Way, Princeton, NJ 08540*

[3]*Department of Physics, University of Wisconsin, Madison, WI 53706*

**Free magnetic moments usually manifest themselves in Curie laws, where weak external magnetic fields produce magnetizations diverging as the reciprocal 1/T of the temperature. For a variety of materials that do not display static magnetism, including doped semiconductors[1] and certain rare earth intermetallics[2], the 1/T law is changed to a power law $T^{-\alpha}$ with $\alpha < 1$. We report here that a considerably simpler material, namely an insulating, magnetic salt can also display such a power law, and show via comparison to specific heat data[3] and numerical simulations that quantum mechanics is crucial for its formation. Two quantum mechanical phenomena are needed, namely level splitting – which affects the spectrum of excited states – and entanglement – where the wavefunction of a system with several degrees of freedom cannot be written as a product of wavefunctions for each degree of freedom. Entanglement effects become visible for remarkably small tunnelling terms, and are turned on well before tunnelling has visible effects on the spectrum. Our work is significant because it illustrates that entanglement is at the heart of a very simple experimental observation for an insulating quantum spin system.**

The insulator we focus on in the search for the cause of the anomalous power law divergence of the magnetic susceptibility is $LiHo_xY_{1-x}F_4$, a salt where magnetic $Ho^{3+}$ ions are randomly substituted for nonmagnetic $Y^{3+}$ with probability $x$. For $x = 1$, the material is the dipolar-coupled ferromagnet, $LiHoF_4$, with a Curie temperature of 1.53 K. Randomly distributing dipoles in a solid matrix provides quenched disorder, while the angular anisotropy of the dipole-dipole interaction leads to competition between ferromagnetic and antiferromagnetic bonds and the possibility of many (nearly) degenerate ground states[4]. Indeed, the low temperature magnetic phase diagram of the dipolar-coupled rare earth tetrafluorides progresses smoothly from long-range order to glassiness with increasing spin dilution[3]. What interests us here, however, is the considerably diluted $x = 0.045$ compound, where we have observed[5,6] – contrary to classical expectations[4]– novel 'antiglass' behaviour as well as long-lived spin oscillations whose qualitative understanding seems to require mesoscopic quantum coherence. We show in Fig. 1 the experimental dc susceptibility plotted against temperature for a single crystal specimen of the material. What emerges is not the standard Curie law 1/T expected for non-interacting magnetic moments, but instead a diverging response following a power law $T^{-\alpha}$, with $\alpha = 0.75\pm0.01$. This power law is close to that associated with the diverging local susceptibilities inferred for doped silicon[1] as well as metallic rare earth materials[2] on the brink of magnetic order. What is most striking, however, is that the magnetic susceptibility for $LiHo_{0.045}Y_{0.955}F_4$ is a smoothly diverging quantity even though the magnetic specific heat (Fig. 2a) is characterized by unusually sharp peaks in the same temperature range. In ordinary materials containing magnetic ions, there is a strong correlation between magnetic susceptibility and specific heat in the sense that anomalies, especially as strong as the sharp peaks in the specific heat, are reflected in the susceptibility.

The data for $LiHo_{0.045}Y_{0.955}F_4$ thus provide three puzzles: the absence of a spin glass transition predicted for a collection of randomly placed dipoles, the anomalous power law behaviour $\chi \sim T^{-\alpha}$, and the coexistence of a featureless power law in $\chi$ with sharp anomalies in the specific heat. We show here that it is an intrinsic quantum mechanical term in the dipole Hamiltonian that stabilizes the spin liquid 'antiglass' and resolves the puzzles. Following a pair-wise decimation procedure[7-13] but adapted to treat the full axial and transverse components of the dipole-dipole interaction, we find that quantum fluctuations continue to provide channels for relaxation down to the lowest temperatures. We simulate the evolution with temperature T of both the magnetic susceptibility and the heat capacity using the actual interaction parameters between Ho moments obtained from various experimental results, and compare quantitatively the results of simulation and experiment.

LiHoF$_4$ crystallizes in a body-centred tetragonal (CaWO$_4$) structure with lattice constants[14] $a = a' = 5.175(5)$ Å and $c = 10.75(1)$ Å. Each unit cell has four formula units with the magnetic Ho$^{3+}$ ions occupying positions (*0, 0, 0*), (*0, a/2, c/4*), (*a/2, a/2, c/2*) and (*a/2, 0, 3c/4*). The Ising axis is defined by the crystal field of the Ho$^{3+}$ ions that forces the spin 1/2 magnetic moments with a g-factor of 13.8 to point along the crystalline *c*-axis. We generate a model of the three dimensional lattice of LiHo$_x$Y$_{1-x}$F$_4$ on the computer by repeated translation of the unit cell vectors. N spins are distributed randomly in this lattice with probability of occupancy *x* on the body-centred tetragonal sites. Periodic boundary conditions are applied. We have used a maximum of N = 400 spins ($8 \times 10^4$ pair-wise interactions) and have checked that our results are in the N-independent limit. The Ising spin ($\sigma_i^z$) at each site i is assigned a value of 1 or $-1$ randomly. We have confirmed explicitly that in the dilute limit of *x* the outcome is not sensitive to this particular initial condition, obviating the need to average over initial spin configurations. Our simulations incorporate pair-wise dipolar couplings in which the interaction energy between two magnetic dipole moments $\vec{M}_1$ and $\vec{M}_2$ separated by the vector $\vec{r} = \hat{r}r$ is

$$E_{int} = \frac{1}{r^3}\left(\vec{M}_1 \cdot \vec{M}_2 - 3(\hat{r} \cdot \vec{M}_1)(\hat{r} \cdot \vec{M}_2)\right). \tag{1}$$

The magnetic dipole moment $\vec{M}_i$ at site i is related to the spin $\vec{\sigma}_i$ via $M_{i\lambda} = \mu_i g_\lambda \sigma_{i\lambda}$ ($\lambda$=x, y, z) where $\mu_i$ is the magnetic moment of the i$^{th}$ spin, initially $\mu_i = \mu_B/2$ for all i, and the elements of the anisotropic g-factor matrix ($g_x=g_y=g_\perp=0.74$, $g_z=g_{ll}=13.8$) are known from previous measurements on the pure material[14-16].

Our first simulation is a classical calculation in which $g_\perp= 0$ and eq.(1) reduces to $-2M_{1z}M_{2z}/r^3$. The Hamiltonian can be written as:

$$H = -\sum_{i,j}^{N} J_{ij}\mu_i \sigma_{iz} \mu_j \sigma_{jz} \qquad (2)$$

where $J_{ij}$ (which assimilates the numerical constants and the product $g_{ll}^2$ ) falls off as $1/r^3$ and the $\sigma$'s are classical Ising spins that can take the values ±1. It is expected to be valid both in the very dense (x = 1) ferromagnetic and very dilute (x → 0) paramagnetic limits. We adopt the nomenclature of "axial dipole" Hamiltonian to describe eq.(2) since the dipolar field of spin j coupling to the moment of spin i acts only along the Ising axis. We show in Fig. 3a the energy levels calculated using eq.(2) for a spin pair (i,j) when $|\mu_i| = |\mu_j|$.

Once we have calculated the energy of all spin pairs, we arrange the pairs in a hierarchy based on their coupling strength and pick the pair with the largest energy $|J_{max}|$. If $2|J_{max}| > k_BT$, the excited state $+J_{max}$ becomes redundant and this pair is forced into its ground state. The pair is replaced by a *composite* single spin of equivalent net magnetic moment that can be either $\mu_C=|\mu_i|-|\mu_j|$ (antiferromagnetic interaction) or $\mu_C=|\mu_i|+|\mu_j|$ (ferromagnetic interaction). If the net moment is zero, the two spins are decimated completely and removed from further consideration; otherwise they are replaced by one spin with the new composite moment placed at the average position of the two spins in the pair. Only the magnetic moment $\mu_C$ and the position $\vec{r}_C$ of the pair are renormalized; $g_{ll}$ is left unchanged. The new magnetic dipole moment $M_{Cz}$ is now given by $\mu_C g_{||}\sigma_{Cz}$ ($M_{Cx}= M_{Cy}= 0$). The altered demography requires the procedure to begin anew with the calculations of the pair-wise interactions between the remaining spins and the composite $\vec{M}_C$ at $\vec{r}_C$ using eq.(2). At each iteration one spin pair is either eliminated or transformed into a single spin degree of freedom until the strongest pair remaining has a gap $2|J_{max}| < k_B T$. All spins remaining in the system are considered "free" at that T. We choose T = 1 K at the outset given that the nearest-neighbour interaction energy $J_{nn}$ = 1.2 K; the temperature is then lowered in steps $\Delta T$ = 0.01 K down to T = 0.01 K.

At each temperature the calculation produces a list $\{M_i\}$ of N(T) 'free' moments. Given that the susceptibility for a free Ising moment $M_i$ is $M_i^2/k_B T$ and its contribution to the entropy is $(R\ln 2)N(T)$, we can compute the specific heat and susceptibility via the relations:

$$\chi = \sum_i \frac{M_i^2}{k_B T} \qquad (3)$$

and

$$C = \frac{TdS}{dT} = RT\ln 2 \frac{dN(T)}{dT} \qquad (4)$$

Fig. 2c reveals one success of the classical decimation, namely the appearance of sharp features in the specific heat. While the features are of roughly the same magnitude as seen in the experiment, they do not occur at the correct temperatures. Fig. 1, where the filled green circles are the results of the classical calculation, reveals a more disturbing problem. In accord with intuition, but in disagreement with experiment, there are sharp anomalies in $\chi$ which coincide with the peaks in C. Moreover, the classical susceptibility at low temperatures is over an order of magnitude smaller than that measured.

The axial dipole Hamiltonian in eq.(2) takes into account only the interactions parallel to the Ising axis by ignoring $g_\perp$ and treats the spins as classical bits rather than Pauli matrices. For pure LiHoF$_4$ in its ferromagnetic state, this is the complete picture because the lattice symmetry ensures that at any site the perpendicular component of the dipolar field due to the other spins sums to zero. However, as the magnetic Ho$^{3+}$ ions are randomly replaced by non-magnetic Y$^{3+}$, all transverse components are no longer perfectly compensated, and at dipole concentrations of only a few percent the site-specific, internal transverse fields can be as large as 1 kOe. With a finite $g_\perp$, the full Hamiltonian, eq.(1), no longer reduces to eq.(2) but acquires off-diagonal terms of the form $\beta\sigma_{ix}\sigma_{jz}$ where $\beta$ includes numerical constants and the product $g_\perp g_\parallel$. Terms of order $\sigma_{ix}\sigma_{jx}$ are not included, because $g_\perp^2 \ll g_\perp g_\parallel \ll g_\parallel^2$.

The results in Figs. 1 and 2 demonstrate how the decimation calculation is affected when the energy levels but not the eigenfunctions are modified to account for the inclusion of the off-diagonal terms of the dipolar interaction. While the basic decimation scheme remains the same, there are now two energy scales available for comparison with temperature: $2|J|$ and $(J^2+\beta^2)^{1/2} - J$ which is $\sim \beta^2/2J$ to first order. Since this second gap is much smaller than the first it becomes clear, that at a temperature T the number of free spins, N(T), is greater than in the classical case (see Fig. 3b). The results of this modification meet with partial success. The increase in N(T) can account better for the specific heat characteristics (Fig. 2b), but not for the susceptibility, especially at low temperatures (Fig 1, filled green circles) implying that it is not merely the excess in the number of free spins that enhances and smoothes the susceptibility of the sample.

The key to matching the experimental susceptibility result is to employ the quantum mechanical expression[17] for $\chi$:

$$\chi = \sum_{i}^{N} \frac{1}{\sum_{n} \rho_n(i)} \sum_{n} \left( \frac{\left(E_n^{(1)}(i)\right)^2}{k_B T} - 2E_n^{(2)}(i) \right) \rho_n(i) \tag{5}$$

with

$$\rho_n(i) = \exp\left(\frac{-E_n^{(0)}(i)}{k_B T}\right) \; ; \; E_n^{(1)}(i) = \langle n(i)|M_{iz}|n(i)\rangle \text{ and } E_n^{(2)}(i) = \sum_{m} \frac{|\langle n(i)|M_{iz}|m(i)\rangle|^2}{E_m^{(0)}(i) - E_n^{(0)}(i)}$$

where first we sum over the energy levels of the $i^{th}$ effective spin remaining, and then sum over all N effective spins. Except for a dwindling population with T of isolated, unrenormalized spins, $|n\rangle$ and $|m\rangle$ are now the entangled pair eigenstates illustrated in Fig. 3b. Note that keeping only the first term in the brackets reduces eq.(5) to eq.(3). The result obtained using eq. (5) (Fig. 1, black circles) agrees quantitatively with the actual measurements (Fig.1, red triangles) of the magnetic susceptibility in the dc limit for $LiHo_{0.045}Y_{0.955}F_4$.

The inclusion of the off-diagonal terms, $E_n^{(2)}$, produces the reconciliation between simulation and experiment by entangling the antiferromagnetic ground state with the ferromagnetic excited state (Fig. 3b). In this manner, the excited classical states 'frozen out' by the decimation still enter into the expression for the susceptibility, thereby enhancing its value. The concurrence[18,19], which ranges from 0 (disentangled states) to 1 (completely entangled), quantifies entanglement in an exact way for bipartite systems. We identify the T-dependent entanglement $\tau$ as the concurrence of the pair wavefunctions contributing to the susceptibility at each T, weighted by the fraction of actual spins involved in the history of the pair formation within the decimation calculation. We find that $\tau = 0.11$ at 0.8 K, growing to 0.88 at 0.01 K, in accord with the trend from near agreement of the 'entangled' and 'quantum level' bulk susceptibilities at 0.8 K towards a factor of four difference at 0.01 K. Moreover, we find that only the slightest degree of entanglement can have profound effects. We illustrate this in Fig. 4, which shows how the two calculations evolve with increasing $g_\perp$. The figure reveals that a $g_\perp/g_\parallel$ as small as $10^{-4}$ produces a large change in $\chi$. The effects of the energy level distribution are minimal by comparison.

Contrary to intuition and common experience, a dilute assembly of Ising dipoles does not freeze when cooled to milliKelvin temperatures. The computer simulations presented here indicate that it is quantum mechanics – the internal magnetic fields transverse to the Ising axis inherent to the dipole-dipole interaction – that stabilizes the spin liquid. However, unlike conventional spin liquids where the dynamics are dominated by a single gap to triplet excitations, the dilute dipoles form a state with a distribution of such gaps, especially well probed by the specific heat, which shows remarkable releases of entropy at certain well-defined temperatures. At the same time, the magnetic susceptibility increases smoothly with decreasing temperature, but at a rate slower than a Curie law[6]. The smoothness is in marked contrast to the highly structured heat capacity, and can only be understood if quantum mechanical mixing – the entanglement of classical ferromagnetic and antiferromagnetic contributions to the wavefunctions – is taken into account. There is a growing realization[20,21,22] that entanglement is a useful concept for understanding quantum magnets, thus unifying two rapidly evolving areas, quantum information theory and quantum magnetism. The discussions to date have focused on one-dimensional magnets and measures of entanglement with clear theoretical meaning but no simple experimental implementation. Our experiments and simulations represent a dramatic illustration of how entanglement, rather than energy level redistribution, can contribute significantly to the simplest of observables – the bulk susceptibility – in an easily stated model problem.

**Acknowledgements**

We are grateful to R. Parthasarathy for illuminating discussions. The work at the University of Chicago was supported by the MRSEC Program of the National Science Foundation, that in Madison by the Petroleum Research Fund and the National Science Foundation, and that in London by a Wolfson-Royal Society Research Merit Award and the Basic Technologies programme of the UK Research Councils.

**Competing interests statement**

The authors declare that they have no competing financial interests.

Correspondence and requests for materials should be addresses to T.F.R (email: t-rosenbaum@uchicago.edu).

FIGURE CAPTIONS

**Figure 1** Magnetic susceptibility $\chi$ vs. temperature T from simulations of the diluted, dipolar-coupled Ising magnet, LiHo$_{0.045}$Y$_{0.955}$F$_4$ compared to experimental data (red triangles). The green circles represent classical decimation when the calculations are performed with $g_\perp$=0. The blue circles represent susceptibility computed using the classical procedure eq.(3) of determining Curie constants by adding (subtracting) moments when the ground state is predominantly ferromagnetic (antiferromagnetic), but with quantum decimation, using energy levels derived from the full dipolar Hamiltonian of eq.(1). While the susceptibility approaches that of the experiment more closely than before, it still deviates by at least a factor of four at low temperatures. The black circles use quantum decimation as well as the correct quantum mechanical form of susceptibility given by eq.(5), utilizing the entanglement of the low-lying energy doublet with the excited states. The line is a best fit to $\chi(T) \propto T^{-\alpha}$, with $\alpha$ = 0.75±0.01. Although $\alpha$

is always less than 1, it is not a universal number. It varies from 0.62 to 0.81 as the concentration $x$ decreases from 0.1 to 0.01, a trend also observed in Heisenberg systems[7]. The simulation results have not been scaled and agree quantitatively with the experimental results.

**Figure 2** Comparison of the temperature-dependent experimental electronic specific heat C(T) for LiHo$_{0.045}$Y$_{0.955}$F$_4$ with different simulation techniques. **a)** Experimental data showing two sharp Schottky-like features dominating the thermal response. **b)** Quantum decimation emphasizing how the well-defined energy levels result in a more complex temperature-dependent specific heat with greater resemblance to the experimental data, especially at low temperatures. Notably, there is the appearance of a sharp peak at 130 mK, close to a similar feature in the data. **c)** Classical decimation demonstrating some success in calculating C(T) but the characteristic sharp features occur at incorrect temperatures. The features occur as $k_B T$ moves through maxima in the distribution of dipolar couplings and are affected as the concentration $x$ varies; this distribution is granular because the dipolar interaction is being sampled between points on a lattice rather than in continuous space.

**Figure 3** Schematic detailing the difference between the classical and the quantum decimation schemes. **a)** The classical energy levels calculated using eq.(2). There are two doubly degenerate energy levels designated $+J$ and $-J$. Depending on the value of the angle formed by the Ising axis and the vector connecting the spins, the ground state $-J$ can correspond to antiferromagnetic or ferromagnetic alignment of ($i,j$). The eigenstates commute with $\sigma_z$ and there is no mixing between the ground state doublet

and the excited states. **b)** The quantum energy levels showing the new entangled eigenstates. $\beta$ is the off-diagonal term of the full dipolar interaction of eq.(1), which reduces to the well-known Hamiltonian of Ising spins in a transverse magnetic field in the limit of small $(g_\perp/g_\parallel)^2$. The two doublets are split to produce eigenstates that are mixtures of the classical states. It is not only states from the same doublet that are mixed: $\beta$ also yields a ground state that mixes ferromagnetic and antiferromagnetic classical states. Thus, the off-diagonal terms in the dipolar interaction introduce both a change in the spectrum, in the form of splittings of the doublets as well as shifts of the 'ferromagnetic' excited states relative to the 'antiferromagnetic' ground state, and mixing, or 'entanglement' of the classical states.

**Figure 4** The change in susceptibility as the quantum entanglement is tuned by varying the ratio of the transverse and longitudinal magnetic g-factors, $g_\perp/g_\parallel$. An arrow denotes the value for $LiHo_{0.0045}Y_{0.955}F_4$. The full quantum susceptibility (filled circles) demonstrates extraordinary sensitivity to the slightest entanglement of the wavefunctions $(g_\perp/g_\parallel \sim 10^{-4})$, while the susceptibility calculated using the quantum energy levels but ignoring the entanglement (open circles) is relatively flat. Quantum entanglement produced by the off-diagonal terms, rather than spectral superposition, dominates the physics.

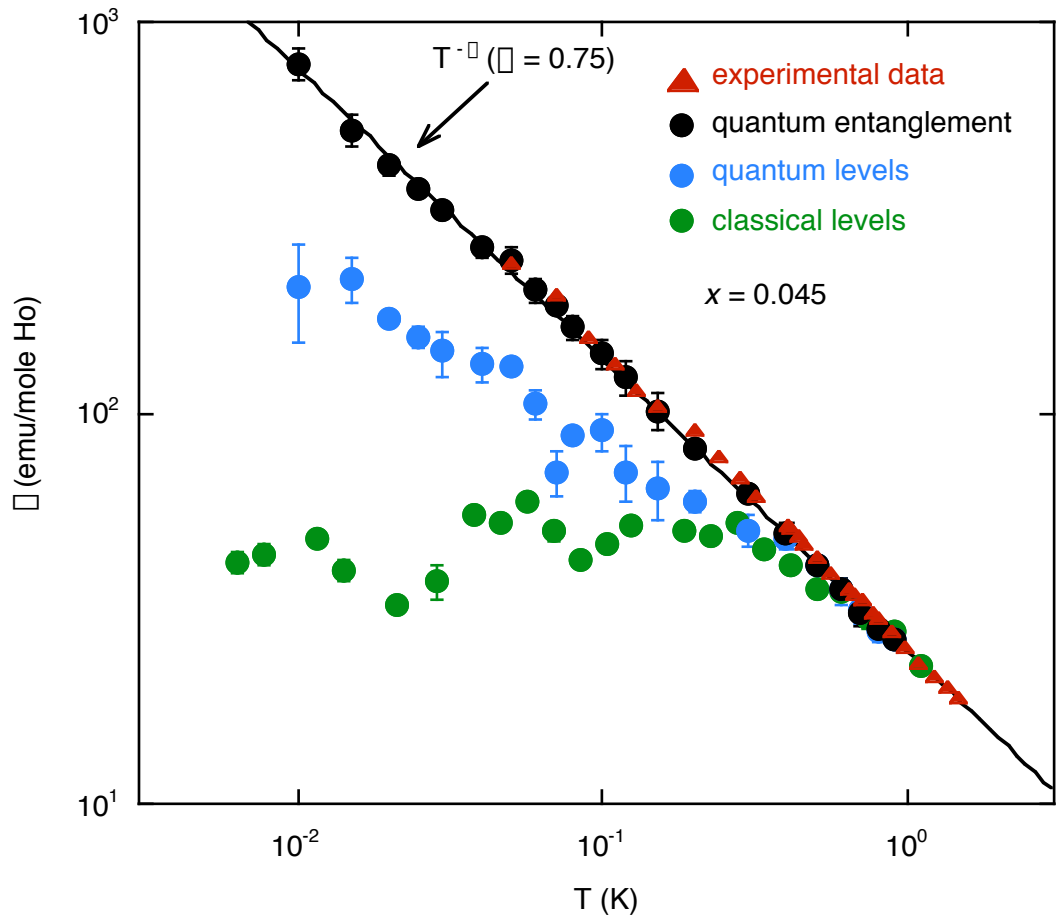

Ghosh_fig1

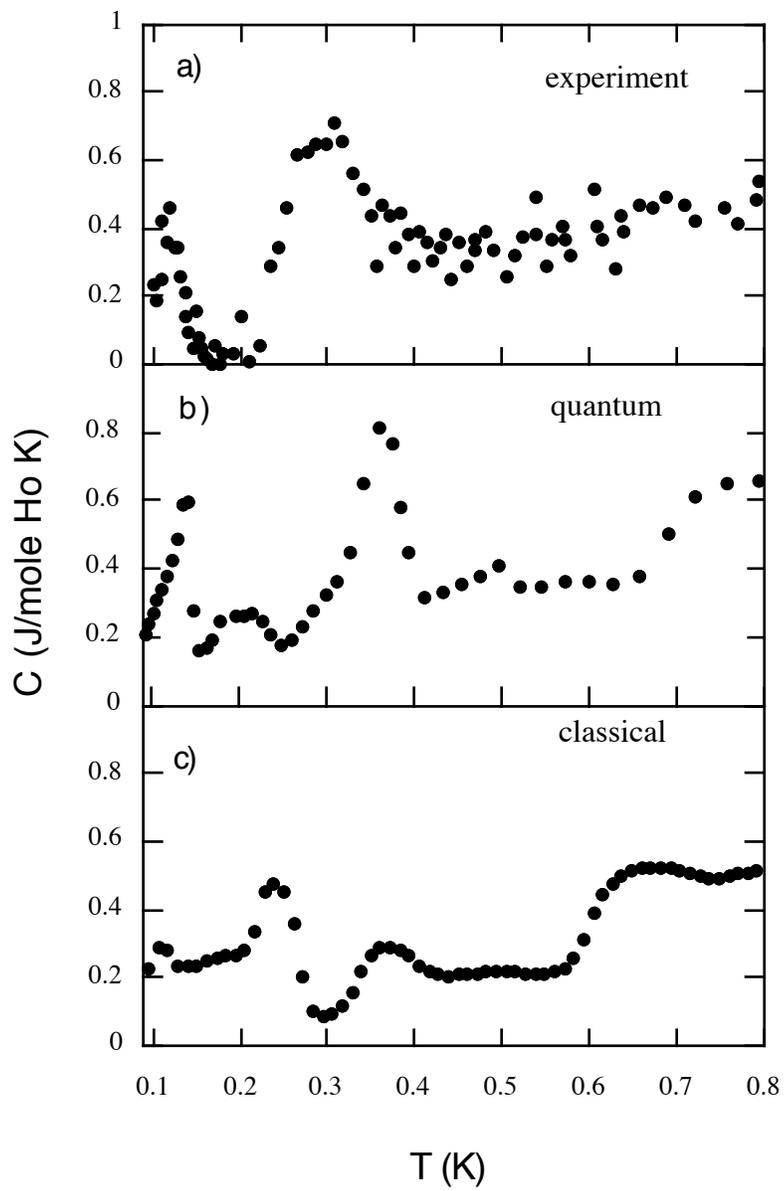

Ghosh_fig2

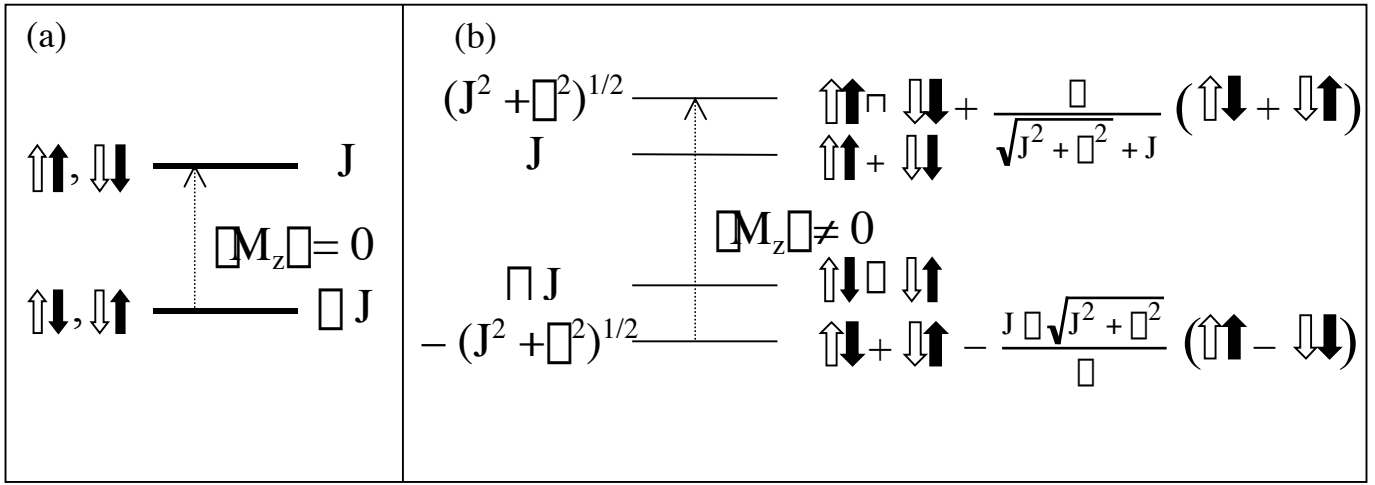

Ghosh_fig3

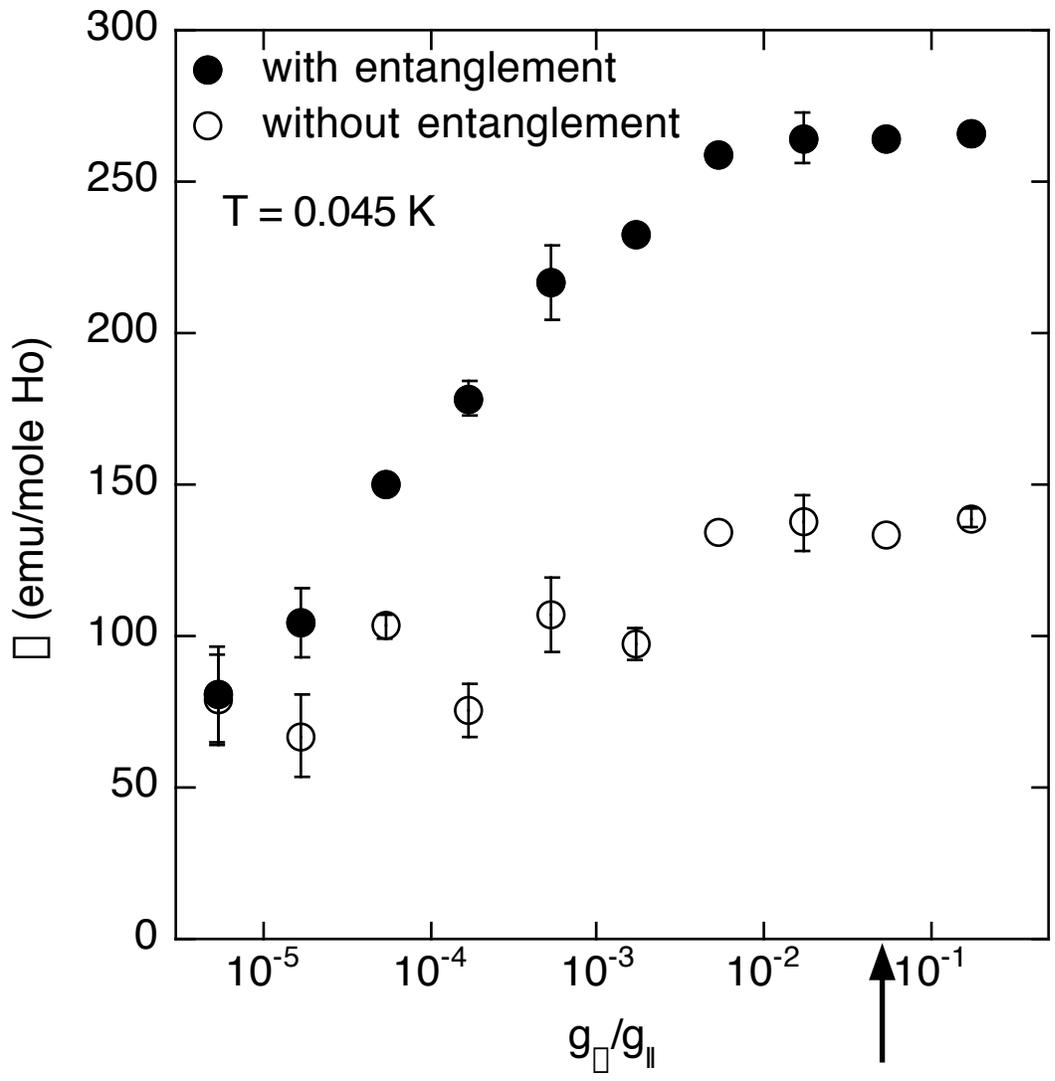

Ghosh_fig4